\documentstyle[epsfig,prl,aps]{revtex}
\begin{document}
\topmargin 0.0in
\preprint{WISC-MILW-99-TH-05,LIGO-P990019-01}
\title{Observational limit on gravitational waves from binary neutron
stars in the Galaxy}
\author{
B.~Allen$^1$,
J.K.~Blackburn$^2$,
P.R.~Brady$^3$,
J.D.E~Creighton$^{1,4}$,
T.~Creighton$^4$,
S.~Droz$^5$,
A.D.~Gillespie$^2$,
S.A.~Hughes$^4$,
S.~Kawamura$^2$,
T.T.~Lyons$^2$,
J.E.~Mason$^2$,
B.J.~Owen$^4$,
F.J.~Raab$^2$,
M.W.~Regehr$^2$,
B.S.~Sathyaprakash$^6$,
R.L.~Savage,~Jr.$^2$,
S.~Whitcomb$^2$, and
A.G.~Wiseman$^1$.}
\address{
${}^{(1)}$Department of Physics, University of Wisconsin - Milwaukee,
PO Box 413, Milwaukee WI 53201.\\
${}^{(2)}$LIGO Project, MS 18-34, California Institute of Technology,
Pasadena, CA 91125.\\
${}^{(3)}$Institute for
Theoretical Physics, University of California, Santa Barbara, CA 93106\\
${}^{(4)}$Theoretical Astrophysics 130-33, California
Institute of Technology, Pasadena, CA 91125.\\ 
${}^{(5)}$Department of Physics, University of Guelph, Guelph, ON N1G
2W1, Canada.\\
${}^{(6)}$Department of Physics and Astronomy, 
UWCC, Post Box 913, Cardiff CF2 3YB, Wales.}
\date{Received 31 March 1999}
\draft
\wideabs{
\maketitle
\abstract
Using optimal matched filtering, we search 25 hours of data from the
LIGO 40-meter prototype laser interferometric gravitational-wave
detector for gravitational-wave chirps emitted by coalescing binary
systems within our Galaxy.  This is the first test of this filtering
technique on real interferometric data.  An upper limit on the rate
$R$ of neutron star binary inspirals in our Galaxy is obtained: with
$90\%$ confidence, $R< 0.5/\mathrm{hour}$.  Similar experiments with
LIGO interferometers will provide constraints on the population of
tight binary neutron star systems in the Universe.
\endabstract
\pacs{PACS numbers: 95.85.Sz, 07.05.Kf, 04.80.Nn, 97.80.-d } 
}
\narrowtext

A world-wide effort is underway to test a fundamental prediction of
physics (the existence of gravitational waves) using a new generation
of gravitational-wave detectors capable of making astrophysical
observations.  These efforts include the US Laser Interferometer
Gravitational-wave Observatory (LIGO) \cite{ligo}, VIRGO
(French/Italian) \cite{others},  GE0-600 (British/German)
\cite{others}, TAMA (Japanese) \cite{others}, and ACIGA (Australian)
\cite{aciga}.  The detectors are laser interferometers with a beam
splitter and mirrors suspended on wires.  A gravitational wave
displaces the mirrors, and shifts the relative optical phase in two
perpendicular paths. This causes a shift in the interference pattern at
the beam splitter~\cite{saulsonbook}.  Within the next decade, these
facilities should be sensitive enough to observe gravitational waves
from astrophysical sources at distances of tens to hundreds of
megaparsecs (Mpc).

During the past 15 years, the LIGO project has used a 40-meter
prototype interferometer at Caltech to develop optical and control
elements for the full scale detectors under construction in Hanford WA
and Livingston LA~\cite{40-meter-refs}.  In 1994, this instrument was
configured as a modulated Fabry-Perot interferometer: light returning
from the two arms was independently sensed~\cite{pla}.  In this
configuration, the detector had its best differential displacement
sensitivity of $\approx 3.5 \times 10^{-19} \text{m}\,\text{Hz}^{-1/2}
$ over a bandwidth of approximately a kHz centered at 600 Hz.

\begin{figure}
\begin{center}
\epsfig{file=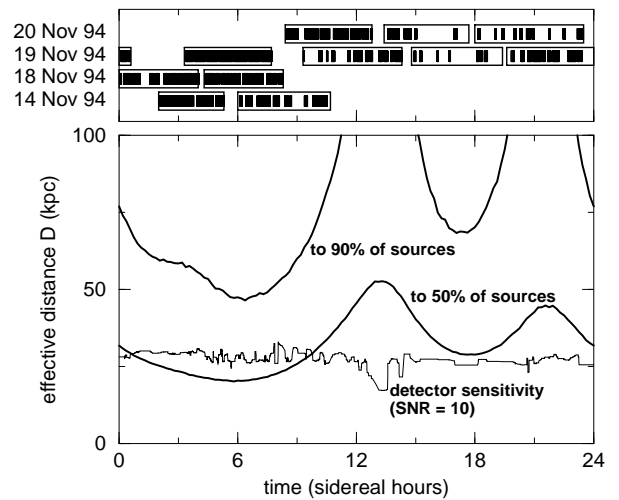,width=8cm}
\caption{\label{f:datatime} 
{\bf Top:} Boxes show data collection times.  Dark bars show data
actually filtered.  {\bf Bottom:} Effective distance $D$ [Eq.~(1)] to
90\% (50\%) of sources varies as the detector antenna pattern sweeps
past the Galactic center.  Dip at 6 hours is when nadir of the
detector (turning with the Earth) points closest to the Galactic
center where the potential sources are clustered.  Fortuitously, much
of the data was taken near such times.  {\bf Jagged Line:} Effective
distance $D$ at which a $2\times 1.4 M_\odot$ optimally oriented
coalescing system would give ${\rm SNR}= \rho = 10$.  This depends on
the average sensitivity of instrument.  The small fluctuations
indicate stable sensitivity.}
\end{center}
\end{figure}

A week-long test run of the instrument was made in November 1994 prior
to a major reconfiguration.  
Fig.~\ref{f:datatime} shows the data-taking periods.
The run yielded 44.8 hours of tape; both arms
were in optical resonance for 39.9 hours (89\% of the time).  Although
the data was taken for diagnostic purposes, it provides an excellent
opportunity to obtain observational limits on
gravitational-wave
sources, and to examine analysis techniques.

A major challenge arises because the real detector noise does not
satisfy the usual simplifying assumptions: stationary and Gaussian.
The 40-m data have the expected colored broad-band background but with
significant deterministic components (spectral peaks), including $\sim
10^2$ sinusoidal components arising from vibration of the support wires
and 60 Hz line harmonics.  There are also transient features occurring
every few minutes: bursts 
\break
of 
noise with durations of $\sim 1 - 500 \> \text{ms}$ 
from accidental 
\noindent
(natural or man-made) disturbances.  These
difficulties led us to develop data analysis techniques that make
matched filtering methods perform well on real data.

This Letter reports on a search of these data for binary inspiral
chirps---the gravitational waveforms produced by pairs of orbiting
stars or black holes.  The search focuses on neutron star binaries in
our Galaxy.  On time scales of $\approx 10^7$ years a binary loses
energy by emitting gravitational waves (primarily at twice the orbital
frequency).  As the orbit shrinks, it circularizes and the period
decreases.  We search for the gravitational waves that would be emitted
during the final few seconds of this process; the stars orbit hundreds
of times per second at separations of tens of km before plunging
together.  (See \cite{40-meter-refs} for results of preliminary
searches.)

The data stream was searched using {\it matched filtering}.  This
method \cite{matched-filtering-lit} uses linear filters constructed
from the expected waveforms, computed using the second post-Newtonian
approximation (2PN) \cite{2ndPNorderwaveforms}.  The 2PN waveform for a
$2\times 1.4 \> M_\odot$ binary is a sweeping sinusoid which enters the
detector pass-band around 120 Hz. The frequency and amplitude increase
during the ensuing 255 cycles; after 1.35 seconds the frequency has
increased to 1822 Hz and the waveform is cut off when the stars merge.
The 2PN approximation results in a reduction of signal to noise ratio
(SNR) $<10\%$ \cite{drozinGRASP}.

The dimensionless strain $h(t)$ of the gravitational wave produces a
differential change $\Delta L(t)=L h(t)$ in the lengths of the two
perpendicular interferometer arms~\cite{saulsonbook}, where $L=38.25
{\rm m} $ is the average arm length.  For a binary system (circular
orbits, no spin) with masses $M=(m_1,m_2)$ this strain is:
\begin{eqnarray}
\label{e:htD}
h(t)\! =\! {1 \> {\rm Mpc} \over D} \!\!\left[ \sin \alpha \> h^{M}_s(t-t_0) 
\!+\!
\cos \alpha \> h^{M}_c(t-t_0) \right] . \nonumber \\
\end{eqnarray}
Here $\alpha$ is a constant determined by the orbital phase and
orientation of the binary system, $t_0$ is the laboratory time when the
chirp signal first enters the detector pass-band, and
$h^{M}_{s,c}(t-t_0)$ are the two polarizations of the gravitational
waveform produced by an inspiraling binary system that is {\it
optimally oriented} at $1\> \textrm{Mpc}$.  If $x$,$y$-axes are defined
by the two interferometer arms then an {\it optimally oriented} binary
system is located on the $z$-axis with its orbital plane parallel to
the $x$-$y$ plane.  The effective distance $D$ depends on the distance
to the source and on its orientation with respect to the detector.  The
detector has a non-uniform response over the sky due to its quadrupolar
antenna pattern.  If the source is not optimally oriented (i.e., not on
the $z$-axis or the orbital plane is tipped), then $D$ is greater than
the source-detector distance.  The formulae for $h^{M}_{s,c}$ are
Eqs.~(2,3a,4a) of Ref.~\cite{2ndPNorderwaveforms}.

The detector signal is the voltage applied to produce a feedback force
on the mirrors to hold the interferometer in resonance; it is
proportional to the differential-displacement $\Delta L(t)$.  This
voltage $v(t)$ was recorded at a sample rate of 9868.42~Hz by a 12~bit
analog-to-digital converter.  Quantizing the data reduces the SNR by
less than {0.9\%}~\cite{AllenBrady}.  The instrument's frequency and
phase response $\tilde R(f)$ was determined at the beginning of each of
eleven $\sim 4$ hour data runs by applying known perturbative forces to
the interferometer ~\cite{sperocalibration}.  These eleven calibration
curves differ by less than $5\%$.  Because errors in calibration affect
the SNR only at second order, we estimate the effects of any
calibration errors or drifts on SNR to be less than $0.3\%$.  The
voltage output $v_h(t)$ that would be produced by a binary inspiral is
given by
\begin{eqnarray}
v_{h}(t) && =  
\int_{-\infty}^t  R(t-t') h(t') dt'  \nonumber \\
&& =  \int_{-\infty}^{\infty}
\tilde h(f) \tilde R^*(f) {\rm e}^{-2 \pi i f t} df \; , \nonumber
\end{eqnarray}
where $Q(t)$ and $\tilde Q(f)$ denote Fourier-transform pairs.

We search for inspiral waveforms using (digital) matched filtering.
Because the inspiral waveforms depend upon the source masses
$M=(m_1,m_2)$ we use a ``bank'' of template waveforms with masses
spaced closely enough \cite{owen} to detect any signal in the mass
range $1.0 \> M_\odot < m_1,m_2 < 3.0 \> M_\odot$ \cite{GRASP}. The
bank contains 687 filters $M_k$ and is designed so that no more than
$2\%$ of SNR would be lost if the mass parameters $M$ of a signal did
not exactly match one of the $M_k$.  For each mass pair $M_k$ in the
template bank two real signals are constructed:
\begin{equation}
\label{e:opfilt}
X^{s,c}_k (t) =  N_k^{s,c} \int_{-\infty}^\infty { \tilde v(f) \tilde
h^{\ast \; M_k}_{s,c} \tilde{R}(f) \over S_v(|f|)} 
{\rm e}^{-2 \pi i f t} \> df.
\end{equation}
These are the outputs of optimal filters matched to the waveform of the
$k$th mass-pair $M_k$.  The denominator $S_v(|f|)$ is (an estimate of)
the one-sided power spectral density of $v(t)$; if the detector's noise
is stationary and Gaussian, then these filters are optimal.  The
normalization factor $N_k^{s,c}$ is chosen so that, in the absence of
any signals, the mean value of $[X^{s,c}_k (t)]^2$ is unity.  We define
the SNR for the $k$th template waveform to be
\[
\rho_k(t) = \mbox{SNR} = \sqrt{\left[ X^s_k (t) \right]^2 + 
\left[ X^c_k (t) \right]^2},
\]
arrived at by maximizing over the phase $\alpha$ of the binary system.
The effective distance $D$ at which coalescence of $2 \times 1.4
M_\odot$ stars would yield an SNR of 10 in the interferometer is shown
in Fig.~\ref{f:datatime}.  (The definition of the SNR follows
Ref.~\cite{owen} and other literature. Its expected value for a source
scales $\propto D^{-1}$.  Its rms value for a single template is
$\sqrt{2}$ in the presence of Gaussian noise alone.)

The data was processed, using FFT methods, in overlapping $\approx
26.6\,\text{s}$ {\it segments} ($2^{18}$ samples). To avoid end
effects, $S^{-1}_v(|f|)$ in Eq.~(\ref{e:opfilt}) was truncated at
$\approx 13.3\,\text{s}$ in the time domain.  The longest chirp signal
was $\approx 2.4\,\text{s}$ long, so the data were overlapped by the
total filter impulse response time of $\approx 15.6\,\text{s}$ ($155
\thinspace 072$ samples) giving a filter output duration of $\approx
10.85\,\text{s/segment}$. Since the process of bringing the optical
cavities into resonance (lock) excites vibrations in the suspension
wires, we discarded the first three minutes of data after each lock
acquisition, allowing the vibrations to damp below other noise
sources.  Of the 39.9 locked hours of data, 8.8 hours were in intervals
too short to analyze; 111 locked intervals remained.  Discarding the
startup transient impulse response of the filters and the first three
minutes of lock yielded $39.9-8.8-6.0 = 25.0$ hours of data analyzed,
in 8289 intervals of filter output (top of Fig.~\ref{f:datatime}).

Poorly understood, non-stationary noise events corrupt the data.
However, these transient events do not have the time-frequency behavior
of inspiral chirps, so we can use the broad-band nature of the
interferometric detector to reject them.  These events are
discriminated from chirps by a $\chi^2$ time-frequency test (Sec.~5.24
of Ref.~\cite{GRASP}).  The frequency band (DC to Nyquist) is divided
into $p$ subintervals, chosen so that for a chirp superposed on
Gaussian noise with the observed power spectrum the expected
contribution to $\rho$ is equal for each subinterval.  One forms a
statistic $\chi^2$ by summing the squares of the deviations of the $p$
signal values from the expected value for the two template
polarizations.  We choose $p=20$ so that Galactic signals that fall at
the maximal template mismatch would not be rejected.  In the presence
of Gaussian noise plus chirp the statistic has a $\chi^2$ distribution
with $2p-2=38$ degrees of freedom~\cite{GRASP}.

Occasionally, there are short sections (i.e., glitches) in the
data when the instrument's output significantly exceeds the rms value.
Some of these glitches were seismically-induced.  These short sections
cause the outputs of the optimal filters to ring, but do not resemble binary
inspiral chirps and are uniformly rejected by the time-frequency
technique described above.  However these glitches bias $S_v(|f|)$
enough to create non-optimal filters. To prevent this problem we
estimate the power spectrum by averaging it for the 8 glitch-free
segments closest in time to the section being analyzed.  The glitches
were identified by seeing if too many samples fell outside a
$\pm3\sigma$ range or any fell outside a $\pm5\sigma$ range.  The
number of segments (8) was chosen to reduce the variance of the
spectrum while still tracking changes in instrument performance.

\begin{figure}
\begin{center}
\epsfig{file=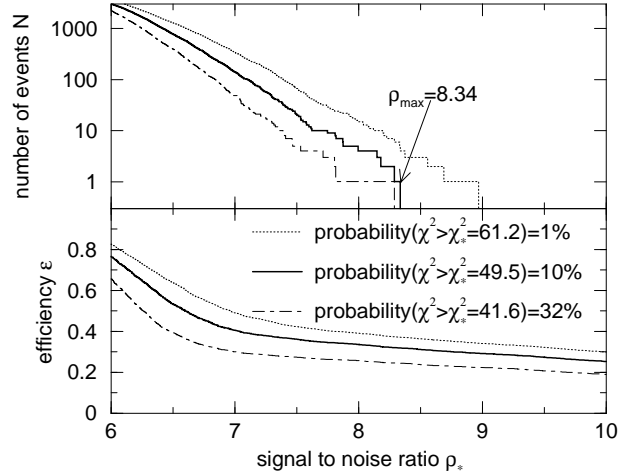,height=2.5in}
\caption{\label{f:eventchi2}
{\bf Top}: total number $N$ of events observed, as a function of the
SNR threshold $\rho_*$ and the threshold $\chi^2_\ast$. {\bf Bottom}:
fraction $\epsilon$ of Galactic inspiral chirp signals that would lie
above SNR threshold $\rho_*$ and below $\chi^2$ threshold $\chi^2_*$.
}
\end{center}
\end{figure}

The data was processed in about 32 hours of clock time on a 48 node
Beowulf computer at UWM (29 Gflops peak). The output of the filtering
process is a list of signals for each segment $j$: the maximum (over
$t$) SNR obtained for each filter $k$ in the bank of 687 filters, the
time $t_j$ at which that maximum occurred, the value of the $\chi^2$
statistic for that filter, and $N^{s,c}$.  In a given segment of data,
we say that an event has occurred if the maximum SNR, over all filters
for which the statistic $\chi^2$ lies below some threshold $\chi^2_*$,
exceeds a threshold $\rho_*$.  The total number $N$ of events observed
in the data set of Fig.~\ref{f:datatime} is plotted as a function of
these thresholds in Fig.~\ref{f:eventchi2}.

Without operating two or more detectors in coinci-
\noindent
dence, it is impossible to characterize the non-Gaussian and
non-stationary background well enough to state with confidence that an
event has been detected.  However one may estimate upper limits on the
rate of Galactic neutron star binary inspirals (Poisson-distributed in
time) using a method which requires minimal assumptions about detector
noise.  Our limit $R_{90\%}$ is based on the probability of a Galactic
neutron star binary signal having an SNR as big as the largest SNR
observed. If the actual inspiral rate is greater than $R_{90\%}$ then
it is likely that we would have observed a larger SNR event.
Fig.~\ref{f:datatime} shows that much of the time the detector was not
pointing at the Galactic bulge; therefore the detector was only
sensitive to a fraction of Galactic binary inspirals.  Thus the
event-rate bound depends on two numbers: (i) the efficiency
$\epsilon_{\text{max}}$ with which the instrument and
filtering/analysis process can detect a binary inspiral in the Galaxy
at the SNR $\rho_{\text{max}}$ of the largest observed event, and (ii)
the total length $T=25.0$ hours of filtered data.

We determined the efficiency $\epsilon$ by Monte-Carlo simulation,
doing additional runs through the data set, and adding simulated
Galactic inspiral waveforms [convolved with the detector response
function $\tilde{R}(f)$] at $30\,\text{s}$ intervals into the detector
output $v(t)$.  {\it This allows us to characterize the detection
process with the properties of the real instrument noise rather than an
ad-hoc model.} The inserted waveforms were drawn from a population of
binary neutron stars with a spatial number distribution given by
$dN\propto e^{-{\cal D}^2/2{\cal D}_0^2}{\cal D}\,d{\cal D}\times
e^{-|Z|/h_Z}dZ$ where ${\cal D}$ is Galactocentric radius, ${\cal
D}_0=4.8$~kpc, $Z$ is height off the Galactic plane, and $h_Z=1$~kpc is
the scale height.  This distribution is similar to the one presented in
Ref.~\cite{CurranLorimer}.  The detection efficiency $\epsilon$ is the
fraction of these simulated inspirals which registered as events in our
filtering/analysis procedure; it increases as the SNR threshold
$\rho_*$ is decreased, or as $\chi^2_*$ is increased, and is shown in
Fig.~\ref{f:eventchi2} for the most-probable mass range \cite{finnmass}
of $1.29$ to $1.45 \> M_\odot$ (the results depend weakly on the
mass).

Our analysis gives an event rate bound.  With $90\%$ confidence, the
rate of binary inspirals in our Galaxy is less than $R_{90\%}=3.89/[T
\epsilon(\rho_{\text{max}},\chi^2_\ast)]$ where
$\rho_{\text{max}}=8.34$ is the largest SNR event observed, and the
threshold $\chi^2_\ast = 49.5$ is chosen so that there is a $10\%$
chance of rejecting a real chirp signal in stationary Gaussian noise.
(This is a Bayesian credible interval computed using a ``uniform prior"
for the event rate.  The dimensionless numerator depends only on the
confidence level.) The efficiency $\epsilon(8.34,49.5)=0.33$ gives
$R_{90\%}=0.5/\mathrm{hour}$.  This is a $90\%$ confidence limit if the
largest event is a real binary inspiral event.  If the largest event is
noise, the confidence is $\ge90\%$.  Thus, $R_{90\%}$ gives a
conservative upper limit on the event rate when the detector noise is
poorly understood. [The on-site environmental monitors show that some
of the larger events in Fig.~\ref{f:eventchi2} arise from seismic
disturbances or laser power fluctuations, but the largest event (on
which our rate limit is based) was detected during normal instrument
operation.]

Let us compare our limit {$R_{90\%} = 0.5/\rm hour$}, with the limit
that could be obtained from the ideal analysis of an instrument that
could detect every Galactic event.  Operating for the same total time
$T=25.0 \> \rm hours$ with an efficiency $\epsilon=1$, the limit
obtained would be three times better:  {$R_{90\%} = 0.17/ \rm hour$}.

Using stellar population models \cite{expected-rates}, 
one can forecast an expected inspiral
rate of $R\sim 10^{-6}\mathrm{ yr}^{-1}$, far below our limit.  
However, unlike these model-based forecasts,
our inspiral limit is based on direct observations of inspirals.  Our
study also demonstrates methods being developed to analyze data from
the next generation of instruments.

A previous search using 100 hours of coincident Glasgow/Garching
interferometer data gave an upper limit on burst sources
\cite{glasgowgarching}.  The current generation of resonant-mass
detectors \cite{resonant} has established upper limits on monochromatic
signals and stochastic background, but neither search addressed the
binary inspiral rate.  A coincidence analysis of bar data for
coalescing binaries might produce a stronger limit than ours.

The full-scale 4-km LIGO interferometers will be much more sensitive
than the 40-meter prototype.  Comprehensive instrument monitoring will
permit detailed characterization of instrument anomalies and removal of
some environmental noise.  Correlation between three independent
instruments will provide lower false alarm rates and greater
statistical confidence.  This will augment the techniques used here and
allow LIGO to detect sources, as well as set tight rate limits.  For
example, if the largest coincident event detected by the LIGO
interferometers has a SNR $\rho_{\mathrm{max}}=5.5$, then we would
obtain the limit
\[
{\cal R}_{90\%} = 6\times 10^{-5}  \>
{\mathrm{Mpc}}^{-3}{\mathrm{yr}}^{-1} \left(
\frac{55{\mathrm{Mpc}}}{r_{{\mathrm{max}}}} \right)^3 \left(\frac{1
{\mathrm{yr}}}{T_{\mathrm{obs}}} \right) \; ,
\]
on the rate of inspiral in the universe, where $T_{\mathrm{obs}}$ is
the observation time, and $r_{\mathrm{max}}$ is the distance to an
optimally oriented source with SNR $\rho_{\mathrm{max}}=5.5$. For the
initial LIGO interferometers, the distance is $r_{\mathrm{max}} = 55 \,
\mathrm{Mpc}$; it will be ten times larger for the enhanced
interferometers, giving an expected rate limit of $ 6\times 10^{-8}
\mathrm{Mpc}^{-3}\mathrm{yr}^{-1}$.  These limits should be compared to
the best guess rate of $8\times10^{-8}
\mathrm{Mpc}^{-3}\mathrm{yr}^{-1}$ given by
Phinney~\cite{expected-rates}.

We thank the LIGO project for making their data and resources
available, and A. Abramovici, R. Spero, and M. Zucker for their
contributions.  P.R.B. thanks the Sherman Fairchild Foundation for
financial support.  J.D.E.C. was supported in part by NSERC of Canada.
B.A. thanks B. Mours and his VIRGO colleagues for their assistance,
and B. Barish, A. Lazzarini, G.  Sanders, K. Thorne, and R.
Weiss for helpful advice.  This work was supported by NSF grants
PHY9210038, PHY9407194, PHY9424337, PHY9507740, PHY9514726,
PHY9603177, PHY9728704, and PHY9900776.  


\vskip -.26in
\begin{thebibliography}{}
\bibitem{ligo}
\vskip -.52in
A.~Abramovici {\it et al}., Science {\bf 256}, 325 (1992).
\bibitem{others}
See B.~Caron {\it et al.}, K.~Danzmann {\it et al.}, and K.~Tsubono in
{\it Gravitational Wave Experiments}, edited by E. Coccia, G. Pizzella,
and F. Ronga, (World Scientific, Singapore, 1995).
\bibitem{aciga}
R.~John~Sandeman, in {\it Second workshop on gravitational wave data
analysis}, edited by M. Davier and P. Hello, ({\'E}ditions
Fronti{\`e}res, Paris, 1998).
\bibitem{saulsonbook}
P.~R. Saulson, {\it Fundamentals of interferometric gravitational wave
detectors}, (World Scientific, Singapore, 1994).
\bibitem{40-meter-refs}
T.~T. Lyons, Ph.D. thesis, Caltech, 1997; A.~D. Gillespie, Ph.D. thesis,
Caltech 1995; S.~Smith, Ph.D. thesis, Caltech, 1988.
\bibitem{pla}
A.~Abramovici {\it et al}., Phys. Lett. A {\bf 218}, 157 (1996).
\bibitem{matched-filtering-lit}
C.~Cutler and {\'E}.~{\'E}. Flanagan, Phys. Rev. D {\bf49}, 2658
(1994); R.~Balasubramanian, B.~S. Sathyaprakash, and S.~V. Dhurandhar,
{\it ibid.} {\bf 53}, 3033 (1996); Erratum {\it ibid.} {\bf54}, 1860
(1996).
\bibitem{2ndPNorderwaveforms}
L.~Blanchet, B.R. Iyer, C.M. Will, and A.G. Wiseman, Class. Quantum
Grav. {\bf 13}, 575 (1996), and references therein.
\bibitem{drozinGRASP}
S.~Droz and E.~Poisson, Sec. 6 in Ref. \cite{GRASP}; S.~Droz, Phys.
Rev. D {\bf 59}, 064030 (1999).
\bibitem{AllenBrady}
B.~Allen and P.~Brady, Report No. LIGO-T970128-01-E, 1997
(unpublished).
\bibitem{sperocalibration}
Robert Spero, Report No. LIGO-T970232-00-R, 1997 (unpublished).
\bibitem{owen}
B.~J. Owen, Phys. Rev. D {\bf53}, 6749 (1996).
\bibitem{GRASP}
B.~Allen {\it et al.}, {\it GRASP: a data analysis package for
gravitational wave detection}, Version 1.8.4.  Manual and package at:
http://www.lsc-group.phys.uwm.edu/.
\bibitem{CurranLorimer}
S.~J. Curran and D.~R. Lorimer, Mon. Not. R. Astron. Soc. {\bf 276},
347 (1995).
\bibitem{finnmass}
L. S. Finn, Phys. Rev. Lett. {\bf 73}, 1878 (1994).
\bibitem{expected-rates}
E.~S. Phinney,  Astrophys. J. {\bf 380}, L17 (1991).
\bibitem{glasgowgarching}
D.~Nicholson {\it et al}., Phys. Lett. A {\bf 218}, 175 (1996).
\bibitem{resonant}
P.~Astone {\it et al}., Astropart. Phys. {\bf 7}, 231 (1997); P.~Astone
{\it et al}., Phys. Lett. B {\bf 385}, 421 (1996); E.~Maucelli {\it et
al}., Phys. Rev. D {\bf 56}, 6081 (1997).
\end{thebibliography}
\end{document}